%% file: conference_101719.tex
\documentclass[conference]{IEEEtran}
\IEEEoverridecommandlockouts
\usepackage{cite}
\usepackage{amsmath,amssymb,amsfonts}
\usepackage{algorithmic}
\usepackage{graphicx}
\usepackage{textcomp}
\usepackage[table]{xcolor}
\usepackage{bm}
\usepackage{bbding}
\usepackage{booktabs}
\usepackage{multicol}
\usepackage{multirow}
\usepackage{hyperref}
\usepackage{amssymb}
\usepackage{colortbl}
\usepackage{svg}
\hypersetup{colorlinks=true, linkcolor=blue, anchorcolor=blue, citecolor=blue}
\def\BibTeX{{\rm B\kern-.05em{\sc i\kern-.025em b}\kern-.08em
    T\kern-.1667em\lower.7ex\hbox{E}\kern-.125emX}}
\begin{document}

\title{
CAS-GAN for Contrast-free Angiography Synthesis
}

\author{
De-Xing Huang$^{1, 2}$,
Xiao-Hu Zhou$^{1, 2, *}$,
Mei-Jiang Gui$^{1, 2}$,
Xiao-Liang Xie$^{1, 2}$,
Shi-Qi Liu$^{1, 2}$,\\
Shuang-Yi Wang$^{1, 2}$,
Hao Li$^{1, 2}$,
Tian-Yu Xiang$^{1, 2}$,
and Zeng-Guang Hou$^{1, 2, *}$,~\IEEEmembership{Fellow,~IEEE} \\
$^1$ State Key Laboratory of Multimodal Artificial Intelligence Systems, \\ Institute of Automation, Chinese Academy of Sciences \\
$^2$ School of Artificial Intelligence, University of Chinese Academy of Sciences \\
{\tt \{\href{mailto:huangdexing2022@ia.ac.cn}{huangdexing2022}, \href{mailto:xiaohu.zhou@ia.ac.cn}{xiaohu.zhou}, \href{mailto:zengguang.hou@ia.ac.cn}{zengguang.hou}\}@ia.ac.cn}
\thanks{*: Corresponding authors.}
}

\maketitle

\begin{abstract}
Iodinated contrast agents are widely utilized in numerous interventional procedures, yet posing substantial health risks to patients. This paper presents {\tt CAS-GAN}, a novel GAN framework that serves as a ``virtual contrast agent" to synthesize X-ray angiographies via disentanglement representation learning and vessel semantic guidance, thereby reducing the reliance on iodinated contrast agents during interventional procedures. Specifically, our approach disentangles X-ray angiographies into background and vessel components, leveraging medical prior knowledge. A specialized predictor then learns to map the interrelationships between these components. Additionally, a vessel semantic-guided generator and a corresponding loss function are introduced to enhance the visual fidelity of generated images. Experimental results on the \textit{XCAD} dataset demonstrate the state-of-the-art performance of our {\tt CAS-GAN}, achieving a FID of $5.87$ and a MMD of $0.016$. These promising results highlight {\tt CAS-GAN}'s potential for clinical applications.
\end{abstract}


\section{Introduction} \label{sec: introduction}
\input{introduction}

\section{Related works} \label{sec: related_works}
\input{related_works}

\section{Methodology} \label{sec: methodology}
\input{methodology}

\section{Experimental setup} \label{sec: experimental_setup}
\input{experimental_setup}

\section{Results} \label{sec: results}
\input{results}

\section{Conclusion} \label{sec: conclusion}
\input{conclusion}

\section*{Acknowledgements}
This work was supported in part by the National Key Research and Development Program of China under 2023YFC2415100, in part by the National Natural Science Foundation of China under Grant 62222316, Grant 62373351, Grant 82327801, Grant 62073325, Grant 62303463, in part by the Chinese Academy of Sciences Project for Young Scientists in Basic Research under Grant No.YSBR-104 and in part by China Postdoctoral Science Foundation under Grant 2024M763535.

\bibliographystyle{IEEEtran}
\bibliography{ref.bib}

\vfill

\end{document}

%% file: introduction.tex
\begin{figure}[htbp]
\centerline{\includegraphics{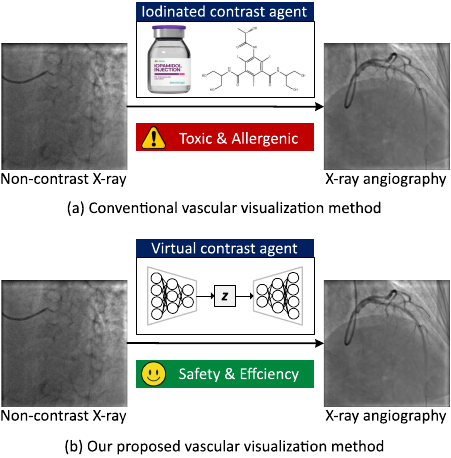}}
\caption{Illustration of two vascular visualization methods.}
\label{fig:story}
\end{figure}

Cardiovascular diseases (CVDs) remain the leading cause of global mortality~\cite{roth2020global}. Image-guided vascular intervention procedures offer minimal trauma and rapid recovery, becoming a mainstream treatment of CVDs~\cite{chandra2021contemporary}. X-ray angiography systems provide physicians with dynamic 2D X-ray images, widely used in interventional procedures. However, since X-rays cannot directly opacify vessels, contrast agents, typically iodine-based, are introduced into vessels to obtain X-ray angiographies~\cite{wagner2021real}, as depicted in Figure~\ref{fig:story} (a). Despite their effectiveness, these contrast agents have several side effects~\cite{manoranjan2020a}, including potentially life-threatening allergic reactions~\cite{clement2018immediate}. Additionally, the elimination of contrast agents via kidneys can exacerbate renal damage, especially in individuals with existing kidney conditions or diabetes~\cite{fahling2017understanding}. Thus, significantly reducing the contrast agent dose or even not using contrast agents while maintaining imaging quality to meet clinical needs is a key challenge that X-ray angiography systems must address~\cite{yin2022precisely}.

Recent advancements in generative adversarial networks (GANs) across both natural and medical imaging domains suggest they can produce convincingly realistic images~\cite{lin2019exploring},~\cite{fan2022tr}. This raises the possibility of \textit{using GANs as “virtual contrast agents” to generate real-looking X-ray angiographies}, as illustrated in Fig.~\ref{fig:story} (b). This will potentially reduce the reliance on iodinated agents and enhance the safety and efficiency of intervention procedures. Specifically, we treat this task as image-to-image translation, \textit{i.e.,} using GANs to generate X-ray angiographies from non-contrast X-ray images.

As one of the most challenging problems in computer vision, image-to-image translation has traditionally been tackled using fully supervised methods like {\tt pix2pix}~\cite{isola2017image}, which require paired images for training. Given the scarcity and high cost of paired medical images, researchers have explored unpaired data methods~\cite{zhu2017unpaired},~\cite{liu2017unsupervised},~\cite{yi2017dualgan}. Cycle-consistency is a simple yet efficient method for learning transformations between source and target domains without paired data and has become a key technique in this field. Yet, existing methods face limitations in clinical application scenarios~\cite{moriakov2020kernel},~\cite{kong2021breaking}. In our task, we necessitate not only the style translation between two image domains but also the precise one-to-one mapping of specific images. For instance, the generated vessels must align with anatomical features in non-contrast X-ray images and maintain continuity. Unfortunately, existing methods often fail to adhere to these constraints, resulting in sub-optimal outcomes.

To tackle these challenges, we introduce {\tt CAS-GAN}, a novel method that learns disentanglement representations to synthesize X-ray angiographies, enhanced by vessel semantic guidance for improved realism. Leveraging medical insights, we assume that anatomical structures between backgrounds and vessels have strong interrelationships. To explicitly formulate such interrelationships, we disentangle backgrounds and vessels in the latent space, using a neural network (termed ``Predictor" in Fig.~\ref{fig:architecture}) to infer vessel representations from background representations. To further ensure the authenticity of anatomical structures in generated vessels, we introduce a vessel semantic-guided generator and a corresponding adversarial loss. The main contributions of this work are summarized as follows:
\begin{itemize}
    \item A novel {\tt CAS-GAN} is proposed for contrast-free X-ray angiography synthesis. To the best of our knowledge, this is the first work in this field.
    \item A disentanglement representation learning method is presented to decouple X-ray angiographies into background and vessel components, and their interrelationships are formulated using neural networks. Furthermore, a vessel semantic-guided generator and a corresponding adversarial loss are introduced to ensure the authenticity of generated images.
    \item Quantitative and qualitative experimental results on the \textit{XCAD}~\cite{ma2021self} demonstrate the state-of-the-art performance of our {\tt CAS-GAN} compared with existing methods.
\end{itemize}

%% file: related_works.tex
\begin{figure*}[t]
\centering
\centerline{\includegraphics{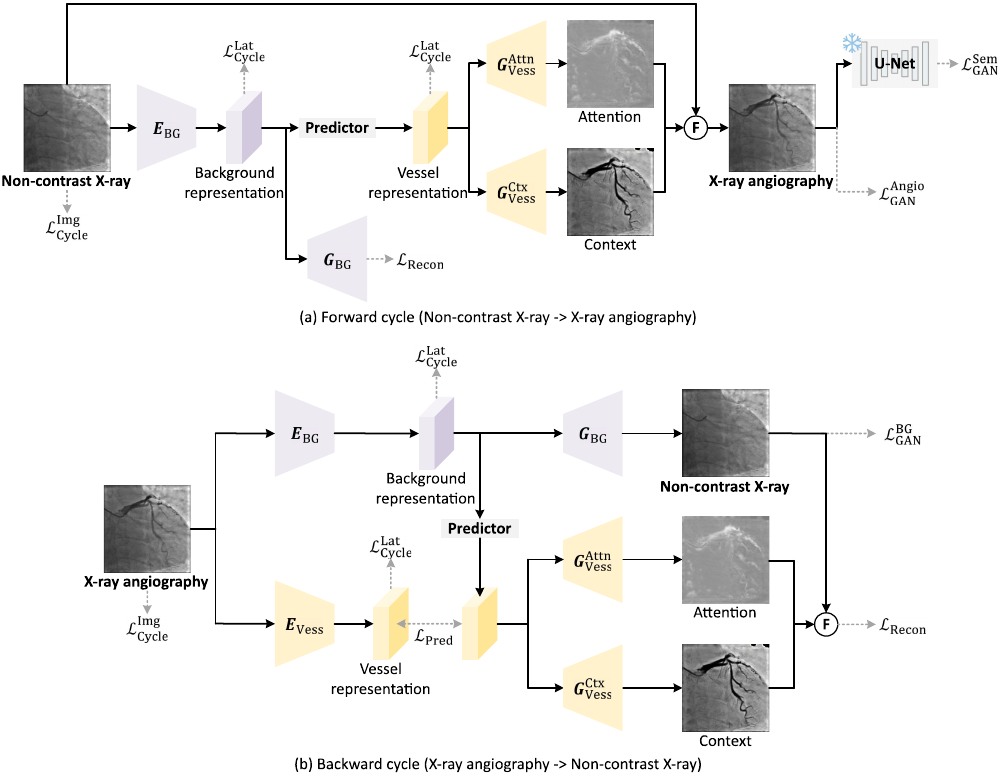}}
\caption{
The overall architecture of our {\tt CAS-GAN}. (a) The forward cycle (Non-contrast X-ray $\rightarrow$ X-ray angiography). (b) The backward cycle (X-ray angiography $\rightarrow$ Non-contrast X-ray). Solid black and dashed gray arrows indicate data flow and loss functions, respectively.
}
\label{fig:architecture}
\end{figure*}

\subsection{Generative adversarial networks}
Generative adversarial networks (GANs), introduced by Goodfellow \textit{et al.}~\cite{goodfellow2014generative} in 2014, revolutionized machine learning by establishing a minimax optimization game between a generator $\bm{G}$ and a discriminator $\bm{D}$. The generator aims to mimic the distribution of real data samples, while the discriminator evaluates their authenticity. Up till now, GANs have been applied to numerous computer vision tasks including medical image generation~\cite{guo2023medgan}, image super-resolution~\cite{park2023content}, and data augmentation~\cite{tran2021data}. Despite their effectiveness, training GANs remains challenging due to issues such as gradient vanishing and mode collapse~\cite{kodali2017convergence},~\cite{tran2018dist},~\cite{bang2021mggan}. Many variants have been developed to address these challenges with more stable training objectives~\cite{salimans2016improved},~\cite{arjovsky2017wasserstein},~\cite{mao2017least}. In our study, we employ GANs to generate X-ray angiographies from non-contrast X-ray images.



\subsection{Image-to-image translation}
Image-to-image translation involves transforming images from one domain to another while maintaining their underlying structures or contents~\cite{choi2020stargan},~\cite{richardson2021encoding},~\cite{martin2023unsupervised},~\cite{armanious2019unsupervised}. Isola ~\textit{et al.}~\cite{isola2017image} pioneered this approach with a conditional GAN framework that utilizes both adversarial~\cite{goodfellow2014generative} and L1 reconstruction losses for training on paired data. However, obtaining a large number of paired data is often impractical, especially in clinical settings. Recent advancements have focused on unpaired data translation, introducing additional constraints like cycle-consistency~\cite{zhu2017unpaired},~\cite{zhou2016learning},~\cite{liu2017unsupervised} to preserve semantic consistency during domain transfer. This constraint ensures that an image converted into another domain and back again should remain largely unchanged, thereby maintaining the original content. Additionally, some studies have incorporated self-supervised contrast loss to enhance semantic relationships between different domains~\cite{park2020contrastive},~\cite{wang2021instance},~\cite{hu2022qs}. 

In the medical imaging field, most works focused on 3D medical imaging modalities, including CT $\rightarrow$ CTA~\cite{lyu2023generative}, MR $\rightarrow$ CT~\cite{kearney2020attention}, image denoising~\cite{armanious2020medgan}, etc. Notably, the most relevant studies to our work are~\cite{martin2023unsupervised},~\cite{tmenova2019cyclegan}. These works involve generating realistic X-ray angiographies from simulated images derived from projections of 3D vessel models. Our research extends these applications by focusing on the generation of vessels in non-contrast X-ray images, presenting a more complex challenge.

\subsection{Disentanglement representation learning}
Disentanglement representation learning is gaining traction for its potential to increase the controllability and interpretability of generated images in image-to-image translation tasks~\cite{chen2016infogan},~\cite{higgins2017beta}. Many works attempt to disentangle images into content and style representations to achieve image translation through swapping style representations~\cite{huang2018multimodal},~\cite{park2020swapping},~\cite{zhang2023breath}. However, these representations are abstract concepts, as they cannot be directly visualized within the image space. Some other works have opted to disentangle images into components with physical significance~\cite{chen2022unpaired},~\cite{pizzati2023physics}, such as separating scenes from elements like raindrops, occlusions, and fog~\cite{pizzati2023physics}. Inspired by these efforts, our work seeks to disentangle X-ray angiographies into background and vessel components and construct the one-to-one mappings of these two components via neural networks.

%% file: methodology.tex
The overall architecture of our method is depicted in Fig.~\ref{fig:architecture}. {\tt CAS-GAN} is designed to learn an unpaired image-to-image translation function, $f:\mathcal{X}\rightarrow\mathcal{Y}$, that maps non-contrast X-ray images $\mathcal{X}$ to X-ray angiographies $\mathcal{Y}$. To address the inherent challenges of this under-constrained mapping, we adopt a cycle-consistency approach similar to {\tt CycleGAN}~\cite{isola2017image}, incorporating an inverse mapping, $g:\mathcal{Y}\rightarrow\mathcal{X}$, to couple with the forward translation. Distinct from conventional image-to-image translation models that primarily focus on style mappings between domains, {\tt CAS-GAN} enhances the process by disentangling backgrounds and vessels from X-ray angiographies. This disentanglement allows us to employ neural networks to explicitly learn the interrelationships between these two distinct components (Section~\ref{sec: dis}). Considering the complex morphology of vessels, which are slender and highly branched~\cite{huang2024spironet}, we further propose a vessel semantic-guided generator (Section~\ref{sec: gen}) and a vessel semantic-guided adversarial loss (Section~\ref{sec: loss}) to enhance the authenticity of generated images.

\subsection{Disentanglement representation learning} \label{sec: dis}
In a specific X-ray angiography $y\in \mathcal{Y}$, it can be conceptualized as a composite of a background (\textit{i.e.,} non-contrast X-ray image $x\in \mathcal{X}$) and vessel components, consistent with its imaging principles~\cite{harrington1982digital}. Physicians can infer vessel morphology from prior anatomical details observed in the backgrounds before injecting contrast agents. Based on this observation, our approach aims to disentangle X-ray angiographies into background and vessel components, subsequently learning a predictor to establish interrelationships between these components.
Specifically, we execute this disentanglement in the latent space using two specialized encoders: a background encoder $\bm{E}_{\rm BG}$ and a vessel encoder $\bm{E}_{\rm Vess}$. Note that we assume the background components of $\mathcal{X}$ and $\mathcal{Y}$ share the same latent space according to the above analysis. The encodings are formulated as follows:
\begin{align}
&z_x^{\rm BG} = \bm{E}_{\rm BG}\left(x\right), z_y^{\rm BG} = \bm{E}_{\rm BG}\left(y\right) \\
&z_y^{\rm Vess} = \bm{E}_{\rm Vess}\left(y\right)
\end{align}
where $z_x^{\rm BG}$, $z_y^{\rm BG}$, and $z_y^{\rm Vess} \in \mathbb{R}^{\frac{C}{r}\times \frac{H}{r}\times \frac{W}{r}}$ are latent space representations. $C$, $H$, and $W$ denote the channel number, height, and width of X-ray images ($x$ and $y$), and $r$ indicates the downsampling ratio of the encoders ($r=4$ in default). Given the absence of $z_x^{\rm Vess}$ in the forward cycle depicted in Fig.~\ref{fig:architecture}, we utilize the paired $\left\{z_y^{\rm BG}, z_y^{\rm Vess}\right\}$ to train the predictor $\bm{M}\left(\cdot\right)$. With the powerful generalization ability of the predictor, the vessel representations $z_x^{\rm Vess}$ of $x$ can be easily derived:
\begin{align}
    z_x^{\rm Vess} = \bm{M}\left(z_x^{\rm BG}\right)
\end{align}
\subsection{Vessel semantic-guided generator} \label{sec: gen}
For the backward cycle, we employ a single generator $\bm{G}_{\rm BG}$ to generate non-contrast X-ray images directly, because this task is comparatively simpler than the forward cycle. The mathematical formulation for the backward cycle $g: \mathcal{Y}\rightarrow \mathcal{X}$ is presented below:
\begin{align}
    x_{\rm g} = \bm{G}_{\rm BG}\left\{\bm{E}_{\rm BG}\left(y\right)\right\} \label{eq: eq1}
\end{align}

The single generator approach is insufficient for the forward cycle, as it cannot discern the most discriminative features of the vessels. Drawing inspiration from attention mechanisms, we introduce a vessel semantic-guided generation to address this limitation. Rather than direct image generation, we deploy two specialized generators: $\bm{G}_{\rm Vess}^{\rm Attn}$ for attention masks $A_g\in \mathbb{R}^{1\times H\times W}$ and $\bm{G}_{\rm Vess}^{\rm Ctx}$ for context masks $C_g\in \mathbb{R}^{C\times H\times W}$. We employ Sigmoid to replace Tanh as the activation function in $\bm{G}_{\rm Vess}^{\rm Attn}$.
\begin{align}
    A_{\rm g} = \bm{G}_{\rm Vess}^{\rm Attn}\left\{\bm{M}\left(z_x^{\rm BG}\right)\right\} \\
    C_{\rm g} = \bm{G}_{\rm Vess}^{\rm Ctx}\left\{\bm{M}\left(z_x^{\rm BG}\right)\right\}
\end{align}

Finally, the corresponding attention masks $A_{\rm g}$, context masks $C_{\rm g}$, and non-contrast X-ray images $x$ are fused to synthesize X-ray angiographies.
\begin{align}
    y_{\rm g} = x\odot \left(1-A_{\rm g}\right) + C_{\rm g}\odot A_{\rm g} \label{eq: eq2}
\end{align}

\subsection{Vessel semantic-guided loss} \label{sec: loss}
The adversarial loss is fundamental in training generative adversarial networks (GANs). However, when two image domains are visually similar, traditional adversarial losses applied directly in the image space may not be sufficiently sensitive. For instance, the primary distinction between non-contrast X-ray images and X-ray angiographies often lies in vascular details, which may not always be distinctly opacified, rendering these images visually akin to one another. To address this, we introduce a vessel semantic-guided adversarial loss, focusing the adversarial training on semantic differences highlighted by vascular information.

Specifically, we employ a pre-trained {\tt U-Net}~\cite{ronneberger2015u} to extract vessel semantic images from both the original $y$ and the generated $y_{\rm g}$ angiographies:
\begin{align}
    &s = {\text{\tt U-Net}}\left(y\right) \\
    &s_{\rm g} = {\text{\tt U-Net}}\left(y_{\rm g}\right)
\end{align}

The vessel semantic-guided adversarial loss is then defined as follows:
\begin{align}
    \mathcal{L}_{\rm GAN}^{\rm Sem} = \mathbb{E}_{s\sim \mathcal{S}}\left\{\log{\bm{D}}_{\rm Sem}\left(s\right)\right\} + \mathbb{E}_{s_{\rm g}\sim \mathcal{S}_{\rm g}}\left\{\log{\left(1-\bm{D}_{\rm Sem}\left(s_{\rm g}\right)\right)}\right\}
\end{align}

This loss function specifically targets the semantic representations of vessels, thereby enhancing the GAN’s ability to discriminate between real and generated images based on vascular features, rather than general imaging characteristics.

\subsection{Training objective}
Due to the highly under-constrained nature of the mappings between the two image domains, incorporating a variety of effective loss functions is crucial for training {\tt CAS-GAN}.

\subsubsection{Prediction loss} We employ the mean square error to train the predictor, enabling it to learn mappings from background to vessel representations:
\begin{align}
    \mathcal{L}_{\rm Pred} = \mathbb{E}_{y\sim \mathcal{Y}} \left\{\left\|\bm{M}\left(z_y^{\rm BG}\right) - z_y^{\rm Vess}\right\|_2^2\right\}
\end{align}

\subsubsection{Adversarial loss}
In addition to the vessel semantic-guided adversarial loss introduced in Section~\ref{sec: loss}, we utilize standard adversarial losses in the image space to train our model:
\begin{align}
    \mathcal{L}_{\rm GAN}^{\rm Angio} &= \mathbb{E}_{y\sim \mathcal{Y}} \left\{\log{\bm{D}_{\rm Angio}\left(y\right)}\right\} \\ \notag &+ \mathbb{E}_{x\sim \mathcal{X}}\left\{\log{\left(1-\bm{D}_{\rm Angio}\left(y_{\rm g}\right)\right)}\right\} \\
    \mathcal{L}_{\rm GAN}^{\rm BG} &= \mathbb{E}_{x\sim \mathcal{X}} \left\{\log{\bm{D}_{\rm BG}\left(x\right)}\right\} \\ \notag &+ \mathbb{E}_{y\sim \mathcal{Y}}\left\{\log{\left(1-\bm{D}_{\rm BG}\left(x_{\rm g}\right)\right)}\right\}
\end{align}
where $x_{\rm g}$ and $y_{\rm g}$ can be represented as Eq.~(\ref{eq: eq1}) and Eq.~(\ref{eq: eq2}).

\subsubsection{Cycle-consistency loss}
Following {\tt CycleGAN}~\cite{isola2017image}, we implement cycle-consistency loss both in the image and latent spaces to ensure the integrity of the transformations:
\begin{equation}
    \begin{aligned}
        \mathcal{L}_{\rm Cycle}^{\rm Img}=\mathbb{E}_{x\sim \mathcal{X}}\left\{\|x_{\rm c}-x\|_1\right\} + \mathbb{E}_{y\sim \mathcal{Y}}\left\{\|y_{\rm c}-y\|_1\right\}
    \end{aligned}
\end{equation}\begin{equation}
    \begin{aligned}
        \mathcal{L}_{\rm Cycle}^{\rm Lat}=\mathbb{E}_{x\sim \mathcal{X}}\left\{\|z_{y_{\rm g}}^{\rm BG}-z_{x}^{\rm BG}\|_1\right\} + \mathbb{E}_{x\sim \mathcal{X}}\left\{\|z_{y_{\rm g}}^{\rm Vess}-z_{x}^{\rm Vess}\|_1\right\} \\ + \mathbb{E}_{y\sim \mathcal{Y}}\left\{\|z_{x_{\rm g}}^{\rm BG}-z_{y}^{\rm BG}\|_1\right\} + \mathbb{E}_{y\sim \mathcal{Y}}\left\{\|z_{x_{\rm g}}^{\rm Vess}-z_{y}^{\rm Vess}\|_1\right\}
    \end{aligned}
\end{equation}
where $x_{\rm c}$ and $y_{\rm c}$ are cycle translation results of $x$ and $y$, respectively.

\subsubsection{Reconstruction loss}
Given our use of autoencoder structures within {\tt CAS-GAN}, the model aims to faithfully reconstruct images post-encoding and decoding:
\begin{align}
    \mathcal{L}_{\rm Recon} = \mathbb{E}_{x\sim \mathcal{X}}\left\{\left\|x_{\rm r}-x\right\|_1\right\} + \mathbb{E}_{y\sim \mathcal{Y}}\left\{\left\|y_{\rm r}-y\right\|_1\right\}
\end{align}
where $x_{\rm r}$ and $y_{\rm r}$ are reconstruction results of $x$ and $y$, respectively.

The training objective for {\tt CAS-GAN} is a weighted sum of the above loss functions, tailored to optimize each aspect of the image translation process:
\begin{align}
    \mathcal{L} &= \mathcal{L}_{\rm GAN}^{\rm Angio} + \mathcal{L}_{\rm GAN}^{\rm BG} + \lambda_1 \mathcal{L}_{\rm GAN}^{\rm Sem} + \lambda_2 \mathcal{L}_{\rm Cycle}^{\rm Img} + \lambda_3 \mathcal{L}_{\rm Cycle}^{\rm Lat} \notag \\ & + \lambda_4 \mathcal{L}_{\rm Pred} + \lambda_5 \mathcal{L}_{\rm Recon}
\end{align}
where $\lambda_1$, $\lambda_2$, $\lambda_3$, $\lambda_4$, and $\lambda_5$ are hyperparameters.

%% file: experimental_setup.tex
\begin{figure*}[t]
\centering
\centerline{\includegraphics{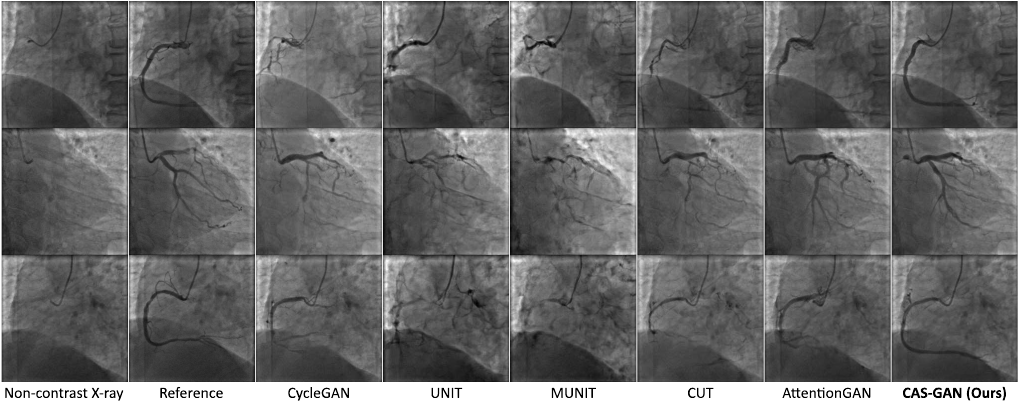}}
\caption{
Qualitative comparisons of X-ray angiographies generated by different models.
}
\label{fig:sota_compare}
\end{figure*}

\subsection{Dataset}
\subsubsection{XCAD~\cite{ma2021self}} 
Our experiments utilize the X-ray angiography coronary vessel segmentation dataset (\textit{XCAD}), which is sourced from a General Electric Innova IGS 520 system. \textit{XCAD} consists of two distinct subsets: \textbf{i)} The first subset includes 1,621 non-contrast X-ray images and 1,621 X-ray angiographies. These images are used for the image translation task; \textbf{ii)} The second subset comprises 126 X-ray angiographies with pixel-wise vessel annotations. This subset is employed to train {\tt U-Net}~\cite{ronneberger2015u} for extracting vessel semantic information. Note that non-contrast X-ray images and X-ray angiographies are not paired, so we serve this problem as an unpaired image translation task. For our experimental setup, we randomly select 621 non-contrast X-ray images and 621 X-ray angiographies as the testing set. The remaining 1,000 non-contrast X-ray images and 1,000 X-ray angiographies are designated as the training set.

\subsection{Implementation details}
We adopt the same generator and discriminator architectures in {\tt CycleGAN}~\cite{isola2017image} for fair comparisons. The predictor within our model is implemented using a multi-layer perceptron (MLP) with a configuration of $C^\prime\rightarrow C^\prime/2\rightarrow C^\prime/4\rightarrow C^\prime/2\rightarrow C^\prime$, where $C^\prime$ is set to $256$ in default. The whole framework is implemented based on PyTorch 2.1.1~\cite{paszke2019pytorch}, Python 3.8.18, and Ubuntu 20.04.6 with an NVIDIA GeForce RTX 4090 GPU. We set hyperparameters of {\tt CAS-GAN} consistent with {\tt CycleGAN}~\cite{zhu2017unpaired}. The input image size is set to $256\times 256$. Adam optimizer~\cite{kingma2014adam} is adopted to train our model for 1,000 epochs. The initial learning rate is set to $2e^{-4}$ and is linearly reduced to zero after 700 epochs. Hyperparameters $\lambda_1$, $\lambda_2$, $\lambda_3$, $\lambda_4$, and $\lambda_5$ are set to $0.5$, $10.0$, $1.0$, $1.0$, and $0.5$.

\subsection{Evaluation metrics}
To evaluate the performance of {\tt CAS-GAN} and other state-of-the-art methods, two popular utilized metrics in medical image translation are selected:

\subsubsection{Fréchet Inception Distance (FID)~\cite{heusel2017gans}}
The Fréchet Inception Distance (FID) measures similarities between two sets of images based on features extracted by {\tt Inception v3}~\cite{szegedy2016rethinking} pre-trained on ImageNet~\cite{deng2009imagenet}. To enhance the feature extraction capability of {\tt Inception v3} for X-ray angiographies, we fine-tuned it on our training dataset.
\input{SOTA_compare}
\subsubsection{Maximum Mean Discrepancy (MMD)~\cite{gretton2012kernel}}
The Maximum Mean Discrepancy (MMD) measures statistical differences between feature distributions of real and generated images. Specifically, image features are extracted by {\tt Inception v3}.

%% file: SOTA_compare.tex
\begin{table}[htbp]
\caption{
Quantitative comparisons with state-of-the-arts on the \textit{XCAD} dataset. The best results are highlighted in \textbf{bold} and the second best results are \underline{underlined}.
}
\label{table:sota_compare}
\centering
\renewcommand\arraystretch{1.2}
\renewcommand\arraystretch{1.2}
\begin{tabular}{lcc}
\toprule
Method & FID $\downarrow$ & MMD ($\times 10$) $\downarrow$ \\ \midrule
{\tt CycleGAN}~\cite{zhu2017unpaired}~{\tiny \color{gray}[ICCV'17]} & $6.54$ & $0.28$ \\
{\tt UNIT}~\cite{liu2017unsupervised}~{\tiny \color{gray}[NeurIPS'17]} & $9.99$ & $\underline{0.22}$ \\
{\tt MUNIT}~\cite{huang2018multimodal}~{\tiny \color{gray}[ECCV'18]} & $8.87$ & $0.33$ \\
{\tt CUT}~\cite{park2020contrastive}~{\tiny \color{gray}[ECCV'20]} & $7.09$ & $0.26$ \\
{\tt AttentionGAN}~\cite{tang2021attentiongan}~{\tiny \color{gray}[TNNLS'21]} & $\underline{6.34}$ & $0.31$ \\
{\tt QS-Attn}~\cite{hu2022qs}~{\tiny \color{gray}[CVPR'22]} & $7.20$ & $0.24$ \\
{\tt StegoGAN}~\cite{wu2024stegogan}~{\tiny \color{gray}[CVPR'24]} & $10.80$ & $2.26$ \\
\rowcolor{red!25} {\tt CAS-GAN}~{\tiny \color{gray}\textbf{[Ours]}} & $\bm{5.87}$ & $\bm{0.16}$ \\
\bottomrule
\end{tabular}
\end{table}

%% file: results.tex
\subsection{Comparisons with state-of-the-arts}
To demonstrate the effectiveness of {\tt CAS-GAN} in X-ray angiography synthesis task, we compared it against several leading unpaired image-to-image translation methods.

Quantitative results on the \textit{XCAD} dataset are presented in Table~\ref{table:sota_compare}. As seen, our {\tt CAS-GAN} outperforms all baselines on FID and MMD. Compared with other methods, {\tt CAS-GAN} and {\tt AttentionGAN} which utilize attention-guided generation can produce more realistic results, reflecting higher FID. However, due to the integration of disentanglement representations and the vessel semantic-guided loss, {\tt CAS-GAN} achieves superior performance compared to {\tt AttentionGAN} on both FID and MMD.
\input{Ablation_1}
Fig.~\ref{fig:sota_compare} showcases qualitative results generated by several baselines and our {\tt CAS-GAN}. Reference images are real X-ray angiographies obtained by injecting contrast agents\footnote{Since reference and non-contrast X-ray images are captured at different moments in the cardiac cycle, quantitative metrics are not applicable for measuring their similarities.}. As seen, baselines frequently exhibit significant entanglement of backgrounds and vessels, leading to highly unrealistic outputs. This issue is particularly evident in the first and third rows of Figure~\ref{fig:sota_compare}, where there is a total loss of structural consistency in vessels. Moreover, baselines fail to infer vessel bifurcations based on background prior. For example, in the second row of Figure~\ref{fig:sota_compare}, all baseline models cannot generate a crucial bifurcation at the position indicated by the catheter. In contrast, {\tt CAS-GAN} can effectively disentangle and understand the complex interrelationships between backgrounds and vessels. This capability allows it to maintain structural integrity and precisely generate vessel bifurcations, markedly outperforming the baselines.

\subsection{Ablation studies}
Extensive ablation experiments are conducted to verify the efficacy of several designs in {\tt CAS-GAN}. Default configurations are highlighted in gray. Quantitative results are detailed in Table~\ref{table:ablation_1}. Our main observations are as follows: \textbf{i)} Learning disentanglement representations facilitates the model's ability to synthesize accurate and realistic vessels based on the given backgrounds, thus leading to $0.11\sim 1.89$ FID improvements; \textbf{ii)} Vessel semantic-guided generator is crucial as it directs the model's focus toward target vessel regions while preserving backgrounds of the original images; \textbf{iii)} Vessel semantic-guided loss can further refine the realism of the generated X-ray angiographies. However, it is observed that integrating VSGL into a baseline model (\textit{i.e.,} Index 1) deteriorates performance. This suggests that VSGL requires the foundational enhancements provided by DRL and VSGG to be effective. Without them, the model struggles to generate high-quality vessels, and VSGL may exacerbate this by trying to force realism in poorly generated images.

\subsection{External validation}
\begin{figure}[htbp]
\centerline{\includegraphics{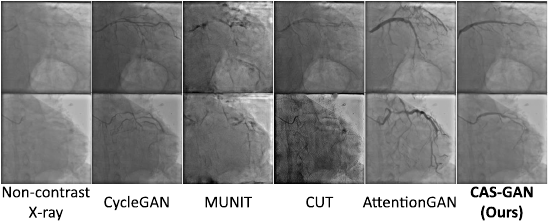}}
\caption{Qualitative comparisons on the external dataset.}
\label{fig:external_compare}
\end{figure}

We also conduct external validations to verify the generalization performance of different models. The data is derived from a collaborating hospital, consisting of 11 X-ray sequences from different patients. From each sequence, we extract two frames (before and after injecting contrast agents) to conduct experiments. Note that we directly evaluate models trained on the \textit{XCAD} dataset without fine-tuning. Fig.~\ref{fig:external_compare} presents several qualitative results. Despite all methods struggling with the significant variance between the \textit{XCAD} and external datasets, {\tt CAS-GAN} showcased relatively superior performance. The specialized design of our model, particularly its disentanglement approach, contributed to its better handling of the dataset's challenges. In contrast, models like {\tt CycleGAN}~\cite{zhu2017unpaired}, {\tt MUNIT}~\cite{huang2018multimodal}, and {\tt CUT}~\cite{park2020contrastive} altered the image appearances significantly, while {\tt AttentionGAN}~\cite{tang2021attentiongan} tended to generate unrealistic vessel structures, entangling with backgrounds. We anticipate that enhancing the dataset size and diversity will significantly improve model performance on external datasets and we will develop a more comprehensive dataset in future works.

%% file: Ablation_1.tex
\begin{table}[htbp]
\caption{
Effects of several designs on the \textit{XCAD} dataset. FID is reported in this table. DRL: Disentanglement representation learning. VSGG: Vessel semantic-guided generator. VSGL: Vessel semantic-guided loss.
}
\label{table:ablation_1}
\centering
\renewcommand\arraystretch{1.2}
\begin{tabular}{llllll}
\toprule
Index & DRL & VSGG & VSGL & FID $\downarrow$ & $\Delta$ \\ \midrule
1 & \textcolor{green}{\XSolidBrush} & \textcolor{green}{\XSolidBrush} & \textcolor{green}{\XSolidBrush} & $7.14$ & ${\color{blue}+1.27}$ \\
2 & \textcolor{green}{\XSolidBrush} & \textcolor{green}{\XSolidBrush} & \textcolor{red}{\Checkmark} & $8.59$ & ${\color{blue}+2.72}$ \\
3 & \textcolor{green}{\XSolidBrush} & \textcolor{red}{\Checkmark} & \textcolor{green}{\XSolidBrush} & $6.57$ & ${\color{blue}+0.70}$ \\
4 & \textcolor{green}{\XSolidBrush} & \textcolor{red}{\Checkmark} & \textcolor{red}{\Checkmark} & $5.98$ & ${\color{blue}+0.11}$ \\
5 & \textcolor{red}{\Checkmark} & \textcolor{green}{\XSolidBrush} & \textcolor{green}{\XSolidBrush} & $6.87$ & ${\color{blue}+1.00}$ \\
6 & \textcolor{red}{\Checkmark} & \textcolor{green}{\XSolidBrush} & \textcolor{red}{\Checkmark} & $6.70$ & ${\color{blue}+0.83}$ \\
7 & \textcolor{red}{\Checkmark} & \textcolor{red}{\Checkmark} & \textcolor{green}{\XSolidBrush} & $5.93$ & ${\color{blue}+0.06}$ \\
\rowcolor{gray!25} 8 & \textcolor{red}{\Checkmark} & \textcolor{red}{\Checkmark} & \textcolor{red}{\Checkmark} & $5.87$ & $-$ \\ \bottomrule
\end{tabular}
\end{table}

%% file: conclusion.tex
This paper introduces {\tt CAS-GAN}, a novel unpaired image-to-image translation model for contrast-free X-ray angiography synthesis. By effectively disentangling X-ray angiographies into background and vessel components within the latent space, {\tt CAS-GAN} can effectively learn the mappings between these components. Moreover, the detailed designed generator and loss function based on vessel semantic guidance can further boost the realism of generated images. Comparative results with other state-of-the-art methods have demonstrated the superior performance of {\tt CAS-GAN}. Our study offers a promising approach to reducing the reliance on iodinated contrast agents during interventional procedures. In future works, we aim to improve {\tt CAS-GAN}'s capabilities by expanding the training dataset and incorporating more advanced models.

%% file: conference_101719.bbl
\begin{thebibliography}{10}
\providecommand{\url}[1]{#1}
\csname url@samestyle\endcsname
\providecommand{\newblock}{\relax}
\providecommand{\bibinfo}[2]{#2}
\providecommand{\BIBentrySTDinterwordspacing}{\spaceskip=0pt\relax}
\providecommand{\BIBentryALTinterwordstretchfactor}{4}
\providecommand{\BIBentryALTinterwordspacing}{\spaceskip=\fontdimen2\font plus
\BIBentryALTinterwordstretchfactor\fontdimen3\font minus \fontdimen4\font\relax}
\providecommand{\BIBforeignlanguage}[2]{{%
\expandafter\ifx\csname l@#1\endcsname\relax
\typeout{** WARNING: IEEEtran.bst: No hyphenation pattern has been}%
\typeout{** loaded for the language `#1'. Using the pattern for}%
\typeout{** the default language instead.}%
\else
\language=\csname l@#1\endcsname
\fi
#2}}
\providecommand{\BIBdecl}{\relax}
\BIBdecl

\bibitem{roth2020global}
G.~A. Roth, G.~A. Mensah, and V.~Fuster, ``The global burden of cardiovascular diseases and risks: A compass for global action,'' \emph{J. Am. Coll. Cardiol.}, vol.~76, no.~25, pp. 2980--2981, 2020.

\bibitem{chandra2021contemporary}
R.~A. Chandra, F.~K. Keane, F.~E. Voncken, and C.~R. Thomas, ``Contemporary radiotherapy: Present and future,'' \emph{Lancet}, vol. 398, no. 10295, pp. 171--184, 2021.

\bibitem{wagner2021real}
M.~G. Wagner \emph{et~al.}, ``Real-time respiratory motion compensated roadmaps for hepatic arterial interventions,'' \emph{Med. Phys.}, vol.~48, no.~10, pp. 5661--5673, 2021.

\bibitem{manoranjan2020a}
M.~S. D’Souza, R.~Miguel, and S.~D. Ray, ``Radiological contrast agents and radiopharmaceuticals,'' \emph{Side Eff. Drugs Annu.}, vol.~42, pp. 459--471, 2020.

\bibitem{clement2018immediate}
O.~Clement \emph{et~al.}, ``Immediate hypersensitivity to contrast agents: The {French} 5-year {CIRTACI} study,'' \emph{Lancet Discov. Sci.}, vol.~1, pp. 51--61, 2018.

\bibitem{fahling2017understanding}
M.~F{\"a}hling, E.~Seeliger, A.~Patzak, and P.~B. Persson, ``Understanding and preventing contrast-induced acute kidney injury,'' \emph{Nat. Rev. Nephrol.}, vol.~13, no.~3, pp. 169--180, 2017.

\bibitem{yin2022precisely}
M.~Yin \emph{et~al.}, ``Precisely translating computed tomography diagnosis accuracy into therapeutic intervention by a carbon-iodine conjugated polymer,'' \emph{Nat. Commun.}, vol.~13, no.~1, p. 2625, 2022.

\bibitem{lin2019exploring}
J.~Lin, Z.~Chen, Y.~Xia, S.~Liu, T.~Qin, and J.~Luo, ``Exploring explicit domain supervision for latent space disentanglement in unpaired image-to-image translation,'' \emph{EEE Trans. Pattern Anal. Mach. Intell.}, vol.~43, no.~4, pp. 1254--1266, 2019.

\bibitem{fan2022tr}
C.-C. Fan \emph{et~al.}, ``{TR-GAN}: Multi-session future {MRI} prediction with temporal recurrent generative adversarial network,'' \emph{IEEE Trans. Med. Imaging}, vol.~41, no.~8, pp. 1925--1937, 2022.

\bibitem{isola2017image}
P.~Isola, J.-Y. Zhu, T.~Zhou, and A.~A. Efros, ``Image-to-image translation with conditional adversarial networks,'' in \emph{Proc. CVPR}, 2017, pp. 1125--1134.

\bibitem{zhu2017unpaired}
J.-Y. Zhu, T.~Park, P.~Isola, and A.~A. Efros, ``Unpaired image-to-image translation using cycle-consistent adversarial networks,'' in \emph{Proc. ICCV}, 2017, pp. 2223--2232.

\bibitem{liu2017unsupervised}
M.-Y. Liu, T.~Breuel, and J.~Kautz, ``Unsupervised image-to-image translation networks,'' in \emph{Proc. NeurIPS}, vol.~30, 2017.

\bibitem{yi2017dualgan}
Z.~Yi, H.~Zhang, P.~Tan, and M.~Gong, ``{DualGAN}: Unsupervised dual learning for image-to-image translation,'' in \emph{Proc. ICCV}, 2017, pp. 2849--2857.

\bibitem{moriakov2020kernel}
N.~Moriakov, J.~Adler, and J.~Teuwen, ``Kernel of {CycleGAN} as a principle homogeneous space,'' in \emph{Proc. ICLR}, 2020.

\bibitem{kong2021breaking}
L.~Kong, C.~Lian, D.~Huang, Z.~Li, Y.~Hu, and Q.~Zhou, ``Breaking the dilemma of medical image-to-image translation,'' in \emph{Proc. NeurIPS}, vol.~34, 2021, pp. 1964--1978.

\bibitem{ma2021self}
Y.~Ma \emph{et~al.}, ``Self-supervised vessel segmentation via adversarial learning,'' in \emph{Proc. ICCV}, 2021, pp. 7536--7545.

\bibitem{goodfellow2014generative}
I.~Goodfellow \emph{et~al.}, ``Generative adversarial nets,'' in \emph{Proc. NeurIPS}, vol.~27, 2014.

\bibitem{guo2023medgan}
K.~Guo \emph{et~al.}, ``{MedGAN}: An adaptive {GAN} approach for medical image generation,'' \emph{Comput. Biol. Med.}, vol. 163, p. 107119, 2023.

\bibitem{park2023content}
J.~Park, S.~Son, and K.~M. Lee, ``Content-aware local gan for photo-realistic super-resolution,'' in \emph{Proc. ICCV}, 2023, pp. 10\,585--10\,594.

\bibitem{tran2021data}
N.-T. Tran, V.-H. Tran, N.-B. Nguyen, T.-K. Nguyen, and N.-M. Cheung, ``On data augmentation for {GAN} training,'' \emph{IEEE Trans. Image Process.}, vol.~30, pp. 1882--1897, 2021.

\bibitem{kodali2017convergence}
N.~Kodali, J.~Abernethy, J.~Hays, and Z.~Kira, ``On convergence and stability of {GANs},'' \emph{arXiv:1705.07215}, 2017.

\bibitem{tran2018dist}
N.-T. Tran, T.-A. Bui, and N.-M. Cheung, ``{Dist-GAN}: An improved {GAN} using distance constraints,'' in \emph{Proc. ECCV}, 2018, pp. 370--385.

\bibitem{bang2021mggan}
D.~Bang and H.~Shim, ``{MGGAN}: Solving mode collapse using manifold-guided training,'' in \emph{Proc. ICCV}, 2021, pp. 2347--2356.

\bibitem{salimans2016improved}
T.~Salimans, I.~Goodfellow, W.~Zaremba, V.~Cheung, A.~Radford, and X.~Chen, ``Improved techniques for training {GANs},'' in \emph{Proc. NeurIPS}, vol.~29, 2016.

\bibitem{arjovsky2017wasserstein}
M.~Arjovsky, S.~Chintala, and L.~Bottou, ``Wasserstein generative adversarial networks,'' in \emph{Proc. ICML}, 2017, pp. 214--223.

\bibitem{mao2017least}
X.~Mao, Q.~Li, H.~Xie, R.~Y. Lau, Z.~Wang, and S.~Paul~Smolley, ``Least squares generative adversarial networks,'' in \emph{Proc. ICCV}, 2017, pp. 2794--2802.

\bibitem{choi2020stargan}
Y.~Choi, Y.~Uh, J.~Yoo, and J.-W. Ha, ``{StarGAN} v2: Diverse image synthesis for multiple domains,'' in \emph{Proc. CVPR}, 2020, pp. 8188--8197.

\bibitem{richardson2021encoding}
E.~Richardson \emph{et~al.}, ``Encoding in style: A {StyleGAN} encoder for image-to-image translation,'' in \emph{Proc. CVPR}, 2021, pp. 2287--2296.

\bibitem{martin2023unsupervised}
R.~Martin, P.~Segars, E.~Samei, J.~Mir{\'o}, and L.~Duong, ``Unsupervised synthesis of realistic coronary artery {X-ray} angiogram,'' \emph{Int. J. Comput. Assisted Radiol. Surg.}, vol.~18, no.~12, pp. 2329--2338, 2023.

\bibitem{armanious2019unsupervised}
K.~Armanious, C.~Jiang, S.~Abdulatif, T.~K{\"u}stner, S.~Gatidis, and B.~Yang, ``Unsupervised medical image translation using cycle-{MedGAN},'' in \emph{Proc. EUSIPCO}, 2019.

\bibitem{zhou2016learning}
T.~Zhou, P.~Krahenbuhl, M.~Aubry, Q.~Huang, and A.~A. Efros, ``Learning dense correspondence via {3D}-guided cycle consistency,'' in \emph{Proc. CVPR}, 2016, pp. 117--126.

\bibitem{park2020contrastive}
T.~Park, A.~A. Efros, R.~Zhang, and J.-Y. Zhu, ``Contrastive learning for unpaired image-to-image translation,'' in \emph{Proc. ECCV}, 2020, pp. 319--345.

\bibitem{wang2021instance}
W.~Wang, W.~Zhou, J.~Bao, D.~Chen, and H.~Li, ``Instance-wise hard negative example generation for contrastive learning in unpaired image-to-image translation,'' in \emph{Proc. ICCV}, 2021, pp. 14\,020--14\,029.

\bibitem{hu2022qs}
X.~Hu, X.~Zhou, Q.~Huang, Z.~Shi, L.~Sun, and Q.~Li, ``{QS-Attn}: Query-selected attention for contrastive learning in {I2I} translation,'' in \emph{Proc. CVPR}, 2022, pp. 18\,291--18\,300.

\bibitem{lyu2023generative}
J.~Lyu \emph{et~al.}, ``Generative adversarial network--based noncontrast {CT} angiography for aorta and carotid arteries,'' \emph{Radiol.}, vol. 309, no.~2, p. e230681, 2023.

\bibitem{kearney2020attention}
V.~Kearney \emph{et~al.}, ``Attention-aware discrimination for {MR}-to-{CT} image translation using cycle-consistent generative adversarial networks,'' \emph{Radiol.: Artif. Intell.}, vol.~2, no.~2, p. e190027, 2020.

\bibitem{armanious2020medgan}
K.~Armanious \emph{et~al.}, ``{MedGAN}: Medical image translation using {GANs},'' \emph{Comput. Med. Imaging Graphics}, vol.~79, p. 101684, 2020.

\bibitem{tmenova2019cyclegan}
O.~Tmenova, R.~Martin, and L.~Duong, ``{CycleGAN} for style transfer in {X}-ray angiography,'' \emph{Int. J. Comput. Assisted Radiol. Surg.}, vol.~14, pp. 1785--1794, 2019.

\bibitem{chen2016infogan}
X.~Chen, Y.~Duan, R.~Houthooft, J.~Schulman, I.~Sutskever, and P.~Abbeel, ``{InfoGAN}: Interpretable representation learning by information maximizing generative adversarial nets,'' in \emph{Proc. NeurIPS}, vol.~29, 2016.

\bibitem{higgins2017beta}
I.~Higgins \emph{et~al.}, ``{{$\beta$}-VAE}: Learning basic visual concepts with a constrained variational framework,'' in \emph{Proc. ICLR}, vol.~3, 2017.

\bibitem{huang2018multimodal}
X.~Huang, M.-Y. Liu, S.~Belongie, and J.~Kautz, ``Multimodal unsupervised image-to-image translation,'' in \emph{Proc. ECCV}, 2018, pp. 172--189.

\bibitem{park2020swapping}
T.~Park \emph{et~al.}, ``Swapping autoencoder for deep image manipulation,'' in \emph{Proc. NeurIPS}, vol.~33, 2020, pp. 7198--7211.

\bibitem{zhang2023breath}
Y.~Zhang, C.~Li, Z.~Dai, L.~Zhong, X.~Wang, and W.~Yang, ``Breath-hold {CBCT}-guided {CBCT}-to-{CT} synthesis via multimodal unsupervised rrepresentation disentanglement learning,'' \emph{IEEE Trans. Med. Imaging}, vol.~42, no.~8, pp. 2313--2324, 2023.

\bibitem{chen2022unpaired}
X.~Chen \emph{et~al.}, ``Unpaired deep image dehazing using contrastive disentanglement learning,'' in \emph{Proc. ECCV}, 2022, pp. 632--648.

\bibitem{pizzati2023physics}
F.~Pizzati, P.~Cerri, and R.~de~Charette, ``Physics-informed guided disentanglement in generative networks,'' \emph{IEEE Trans. Pattern Anal. Mach. Intell.}, vol.~45, no.~8, pp. 10\,300--10\,316, 2023.

\bibitem{huang2024spironet}
D.-X. Huang \emph{et~al.}, ``{SPIRONet}: Spatial-frequency learning and topological channel interaction network for vessel segmentation,'' \emph{arXiv:2406.19749}, 2024.

\bibitem{harrington1982digital}
D.~P. Harrington, L.~M. Boxt, and P.~D. Murray, ``Digital subtraction angiography: Overview of technical principles,'' \emph{Am. J. Roentgenol.}, vol. 139, no.~4, pp. 781--786, 1982.

\bibitem{ronneberger2015u}
O.~Ronneberger, P.~Fischer, and T.~Brox, ``{U-Net}: Convolutional networks for biomedical image segmentation,'' in \emph{Proc. MICCAI}, 2015, pp. 234--241.

\bibitem{paszke2019pytorch}
A.~Paszke \emph{et~al.}, ``{PyTorch}: An imperative style, high-performance deep learning library,'' in \emph{Proc. NeurIPS}, vol.~32, 2019.

\bibitem{kingma2014adam}
D.~P. Kingma and J.~Ba, ``Adam: A method for stochastic optimization,'' in \emph{Proc. ICLR}, 2014.

\bibitem{heusel2017gans}
M.~Heusel, H.~Ramsauer, T.~Unterthiner, B.~Nessler, and S.~Hochreiter, ``{GANs} trained by a two time-scale update rule converge to a local {Nash} equilibrium,'' in \emph{Proc. NeurIPS}, vol.~30, 2017.

\bibitem{szegedy2016rethinking}
C.~Szegedy, V.~Vanhoucke, S.~Ioffe, J.~Shlens, and Z.~Wojna, ``Rethinking the {Inception} architecture for computer vision,'' in \emph{Proc. CVPR}, 2016, pp. 2818--2826.

\bibitem{deng2009imagenet}
J.~Deng, W.~Dong, R.~Socher, L.-J. Li, K.~Li, and L.~Fei-Fei, ``{ImageNet}: A large-scale hierarchical image database,'' in \emph{Proc. CVPR}, 2009, pp. 248--255.

\bibitem{tang2021attentiongan}
H.~Tang, H.~Liu, D.~Xu, P.~H. Torr, and N.~Sebe, ``{AttentionGAN}: Unpaired image-to-image translation using attention-guided generative adversarial networks,'' \emph{IEEE Trans. Neural Networks Learn. Syst.}, vol.~34, no.~4, pp. 1972--1987, 2021.

\bibitem{wu2024stegogan}
S.~Wu \emph{et~al.}, ``{StegoGAN}: Leveraging steganography for non-bijective image-to-image translation,'' in \emph{Proc. CVPR}, 2024, pp. 7922--7931.

\bibitem{gretton2012kernel}
A.~Gretton, K.~M. Borgwardt, M.~J. Rasch, B.~Sch{\"o}lkopf, and A.~Smola, ``A kernel two-sample test,'' \emph{J. Mach. Learn. Res.}, vol.~13, no.~1, pp. 723--773, 2012.

\end{thebibliography}
